\documentclass{aa}
\usepackage{color}
\usepackage{graphicx}
\usepackage{natbib}
\usepackage{txfonts}
\begin{document}

\title{The effects of stellar winds of fast-rotating massive stars in the earliest phases of the chemical 
enrichment of the Galaxy}

\author {G. Cescutti\inst{1}
\thanks {email to: cescutti@oats.inaf.it}
\and    C.Chiappini\inst{2, 3}
\thanks {email to: cristina.chiappini@unige.ch}}

\institute{ Sezione di Astronomia, Dipartimento di Fisica, Universit\'a di Trieste, via G.B. Tiepolo 11, I-34131
\and Geneva Observatory, Ch. des Maillettes 51, 1290 Sauverny, Switzerland
\and INAF - Osservatorio Astronomico di Trieste, via G.B. Tiepolo 11, I-34131}

\date{Received xxxx / Accepted xxxx}

\titlerunning{The effects of stellar winds of fast rotating massive stars }

\authorrunning{Cescutti \& Chiappini }

\abstract
{} {We use the growing data sets of very-metal-poor  stars to
study the impact of stellar winds of fast rotating massive stars on
the chemical enrichment of the early Galaxy.}  {We use an
inhomogeneous chemical evolution model for the Galactic halo to
predict both the mean trend and scatter of C/O and N/O. In one set of
models, we assume that massive stars enrich the interstellar medium
during both the stellar wind and supernovae phases. In the
second set, we consider that in the earliest phases (Z $<10^{-8}$),
stars with masses above 40 $\mathrm{M}_{\odot}$ only enrich the interstellar medium 
via stellar winds, collapsing directly into black holes.}  
{We predict a larger scatter in the C/O and N/O ratios at low metallicities when allowing
the more massive fast-rotating stars to contribute to the chemical enrichment  only via
stellar winds. The latter assumption, combined with the
stochasticity in the star formation process in the primordial Galactic
halo can explain the wide spread observed in the N/O and C/O ratios
in normal very-metal-poor stars.}  
{For chemical elements with stellar yields that depend strongly 
on initial mass (and rotation) such as C, N, and neutron capture elements, within the range of massive stars, 
a large scatter is expected once the stochastic enrichment of the early interstellar medium is taken into account.
We also find that stellar winds of fast rotators mixed with interstellar medium gas are not enough to explain the large 
CNO enhancements found in most of the carbon-enhanced very-metal-poor stars.
In particular, this is the case of the most metal-poor star known to date, HE~1327$-$2326,
for which our models 
predict lower N enhancements than observed 
when assuming a mixture of stellar winds and interstellar medium.
 We suggest that these carbon-enhanced very metal-poor stars
were formed from almost pure stellar wind
material, without dilution with the pristine interstellar medium.}


\maketitle

\section{Introduction}

The very metal-poor stars of the halo play a fundamental
role in chemical evolution since, at metallicities below [Fe/H] $\approx -$2.5, only
type II supernovae have had time to contribute to the interstellar
medium (ISM) enrichment from which these stars formed, thus
offering a way of constraining the nucleosynthesis in massive
stars at low metallicities (e.g. Chiappini et al. 2005, Fran\c cois et al. 2004).

 At present, several thousand of very metal-poor stars
(hereinafter VMP), i.e. stars with metallicities below [Fe/H]$=-$2.
is known (e.g. Cayrel et al. 2004, Christlieb \& Beers 2005; Masseron
et al. 2009). The samples of VMP stars are expected to increase by about 
one order of magnitude in the next years thanks to surveys such as SDSS/SEGUE-2
and LAMOST (see Beers 2010). Larger samples of VMP stars with high-quality
abundances measurements will play a fundamental role in constraining
the stellar nucleosynthesis of the first generations of massive
stars. 

To date, around 20\% of the VMP stars observed show unexpected large C and
N enhancements with respect to solar. 
These stars are called carbon-enhanced metal-poor stars (CEMP). 
The VMP stars not showing such large C and N enhancements are 
called {\it normal} VMP stars (Cayrel et al. 2004, Spite et al. 2005, 2006).
In the case of the two most iron-poor stars known to date, 
these overabundances can reach 100 to 10000 times the ratios 
found in the Sun (Frebel et al. 2005, 2008, 
Christlieb et al. 2002, Norris et al. 2007). 

The CEMP stars are classified according to the presence or absence of s-
and r-process elements. The peculiar abundance of those showing
over-abundances of s-process elements are usually interpreted as caused 
by accretion from an asymptotic giant branch (AGB) companion
(e.g. Masseron et al 2009 and references therein). However, those
without s-process element enhancements were most probably formed from
enriched gas ejected by earlier generations of massive stars
(CEMP-no).

The study of the chemical enrichment of the pristine Universe in CNO,
the most abundant metals, is of particular interest. The CNO chemical
enrichment could have had a important impact in shaping the early IMF
as well as on the production of Li, Be, and B from spallation of C, N,
and O atoms in the early Universe. Moreover, the existence of CEMP-no
stars suggests the synthesis of CNO to have played an important role in
the first stellar generations.

In standard nucleosynthesis (without rotation), N is
essentially a secondary element in massive stars, and has both primary
and secondary components in low- and intermediate-mass stars. Carbon
is produced as a primary element both in massive and
low-and-intermediate mass stars. Oxygen is synthesized as a primary
element in massive stars. Most massive stars would end their lives as
type II SNe, dispersing their chemical make-up into the interstellar
medium (ISM). However, it is believed that the most massive stars
would collapse directly into black holes without the ejection of a
supernova, especially at very-low metallicities (Heger et
al. 2003)\footnote{Another possibility is the formation of faint
supernovae, with large quantities of mixing and fall back as suggested
by Nomoto and co-authors (e.g. Iwamoto et al. 2005)}. In this case,
massive stars would contribute to the chemical enrichment of the ISM
only via stellar winds.

The stellar yields of CNO can be affected by mass loss
and rotation\footnote{Fast-rotating massive stars can
trigger stellar winds even at very-low metallicities -- see Meynet et
al. (2008).}. In particular, the C and N yields of
intermediate-mass stars still suffer from severe uncertainties related not
only to mass loss rates, but also to the difficulties in
modeling transport mechanisms (such as dredge-up episodes). In fact,
there is evidence of additional mixing processes not included in
standard models, triggered by rotation, gravity waves, and thermohaline
mixing (e.g. Charbonnel \& Talon 2005) in this mass range.

 The measurements of C, N, and the C-isotopic ratio in
normal VMP stars (Spite et al. 2005, 2006) have already set strong
constraints on the nucleosynthesis of these elements by the first
generations of massive stars.
In fact, until 2004 
no conclusive data were available for nitrogen in metal-poor halo
stars.   This situation has improved with the First
Stars ESO large program (Cayrel et al. 2004), which obtained
 CNO abundances  for the first time for a sample of giants with metallicities
below [Fe/H]$=-$2.5 (Spite et al. 2005, 2006 -- see also Fabbian et
al. 2009). It turned out that these VMP halo stars had N/O ratios
around solar, suggestive of high levels of production of primary
nitrogen in massive stars. Moreover a large true scatter (more than
the uncertainties in the derived abundances) has been found for the
measured N/O. A large scatter has been also found for neutron capture 
 elements in the same stars. These findings contrast
with the results for alpha elements, which instead presented striking
homogeneous [$\alpha$/Fe] ratios.

In Chiappini et al. (2005 -- hereinafter C05), the implications of the
new CNO data from Spite et al. (2005) on our understanding of nitrogen
enrichment in the Milky Way were investigated. By the time the latter
paper was published, there was no set of stellar yields able to explain
the very metal-poor data from Spite et al. (2005).  C05
concluded that the only way to account for the new data was to assume
that massive stars at low metallicity rotate fast enough to
produce larger amounts of nitrogen. As shown by Meynet \& Maeder
(2002), rotationally-induced mixing transports the C and O produced in
the He-burning core into the H-burning shell, where they are
transformed in primary $^{13}$C and $^{14}$N. The efficiency of this
process increases when the initial mass and rotational velocity
increase, and the metallicity decreases (but see Ekstr\"om et
al. 2008). C05 predicted that massive stars born with metallicities
below Z$ = 10^{-5}$ should produce a factor of 10 up to a few times
100 more nitrogen (depending on the stellar mass) than the values
given by Meynet \& Maeder (2002) for Z $= 10^{-5}$ and for
v$_{\mathrm{ini}}^{\mathrm{rot}} =$ 300 km s$^{-1}$. This prediction has been 
confirmed by subsequent stellar evolution models computed at very low metallicities
(Hirschi 2007), and higher rotational velocities, which were found to
produce much more N than in the models of Meynet \& Maeder
(2002). Chemical evolution models computed with the new stellar
evolution predictions turned out to not only account for the high N/O
in {\it normal}-VMP stars (Spite et al. 2005), but also for the C/O
upturn and low $^{12}$C/$^{13}$C ratios (Spite et al. 2006) at very
low metallicities (Chiappini et al. 2006, 2008). However, 
the same models cannot account for the huge CNO enhancements observed
in CEMP stars.

Finally, C05 suggested that if the stellar yields of N
are strongly dependent on the rotational velocities, hence on the
mass of very-metal-poor massive stars, it is possible to understand
the apparently contradictory finding by Spite et al. (2005) of a large
scatter in N/O and the almost complete lack of scatter in
[$\alpha$/Fe] ratios found in the same very metal-poor halo stars
(Cayrel et al. 2004). Although the observed scatter could be related
to the distribution of stellar rotational velocities as a function of
metallicity, that the neutron capture elements in the same
stars also show a large scatter pointed to a strong variation in the
stellar yields with the mass range of the stars responsible by the
synthesis of these elements.
Cescutti (2008) explains simultaneously the observed spread in the neutron capture 
 elements and the lack of scatter in the alpha elements
as being caused by the stochasticity in the formation of massive stars,
combined with the fact that massive stars of different mass ranges are
responsible for the synthesis of the different chemical elements,
namely: only massive stars with masses between 12 and 30 M$_{\odot}$
contribute to the neutron capture elements, whereas the whole mass
range of massive stars (10 to 80 M$_{\odot}$) contribute to the
production of alpha elements.

 In the present paper we study the CNO evolution in the
early phases of the galactic formation by means of the inhomogeneous
code developed by Cescutti (2008) with the aim of reproducing not only
the mean trend in the chemical abundances of CNO as in Chiappini
et al. (2006), but also the spread in these particular elements. We assume the
spread to be created by the stochasticity in the formation of stars,
as in Cescutti (2008). Our aim is to see whether the same process as
invoked by Cescutti (2008) to explain the scatter observed in the r- and s- process
elements in normal stars also applies to CNO for which a large scatter is also observed. 
We tested the hypothesis that in the case of CNO the scatter is created by the fact that the  
most massive stars in the early Universe (Z $<10^{-8}$ and M $>$ 40 M$_{\odot}$) were fast rotators,
which collapse directly into a black hole (e.g. Heger et al. 2003), but contribute to the chemical enrichment of the 
ISM via stellar winds (Hirschi 2007, Meynet et
al. 2006). We also investigate whether the predicted scatter is enough to account for
the existence of CEMP-no stars.

In Sect. 2 we introduce the observational data we have adopted in the present paper.
In Sect.3 we briefly present our inhomogeneous model for the
Galactic halo. The adopted stellar nucleosynthesis is described in Sect. 4.
Section 5 is devoted to comparing our model predictions to the
 new data set of metal-poor stars in the solar vicinity. Finally, a
discussion of our results and our conclusions are presented in
Sect. 6.

\section{Observational data}

We compare our theoretical predictions with the measured abundances
in halo stars. We distinguish the halo stars in four classes according
to their metallicities: the ultra metal-poor stars (UMP), [Fe/H]$<-4.0$;
the extremely metal-poor stars (EMP), [Fe/H]$<-3.0$; the metal-poor stars [Fe/H]$<-1.0$;
 and the CEMP stars, [Fe/H]$<-1.0$ and [C/Fe]$>0.9$. 
Christlieb et al. (2002, 2004) measured the first UMP star with an [Fe/H]=$-$5.3: HE~0107$-$5240. 
In Frebel et al (2005) and Aoki et al. (2006), we find the analysis of the chemical enrichment
present in HE~1327$-$2326,  the most iron-poor star known to date with [Fe/H]=$-$5.5.
Frebel et al. (2006) show the first determination from a newly obtained VLT/UVES spectrum
of the oxygen abundance in HE~1327$-$2326.
For these two stars, we used  the determination of the CNO abundance with  
the correction to the 1D LTE abundances for 3D effects
taken from Frebel et al. (2008) for HE~1327$-$2326 and Collet et al. (2006) for HE~0107$-$5240.
Norris et al. (2007) measured the abundance of HE~0557$-$8402, which has an iron abundance of
 [Fe/H]=$-$4.75. For this star a measurement of the abundance of carbon is available, but only upper
limits for N and O, so we prefer  not to  use their data,
except in the [C/Fe] vs [Fe/H] plot.

We employed the data from Spite et al. (2005, 2006) for the chemical abundances of C and N,
selecting only the unmixed extremely metal-poor stars of the Galactic halo.
These data are part of the Large Program ``First Stars''.
 For O, we took the results obtained  by Cayrel et al. (2004) into account
with the correction for stellar surface inhomogeneities 
from Nissen et al. (2002); with this correction,
the ratio is smaller in 3-D than  in the 1-D computations. 
We also used the chemical abundances of 9 EMP stars from 
the recent work by Lai et al. (2008), who analyzed a total number of  28 stars.
We used  only the stars for which the abundances for C,
 N, and O were measured.

For the metal-poor stars of the halo, we exploited the recent data
of Fabbian et al. (2008), assuming  non-LTE for the lines
of O and C with an efficiency equal to 1 for the collisions 
with neutral H ($S_{H}$).
Unfortunately, they do not take the N abundance into account:
 for this element in halo stars we 
used the data analyzed by Israelian et al. (2004) for (N/Fe)
 and N/O. 

We also show the data for CEMP stars collected by  Masseron et al. (2009). From their large data set of 
CEMP stars, we adopted only the data for stars with measured  
abundances of C, N, and O simultaneously. Moreover, we 
only considered stars with [Ba/Fe]$ < 1$ (CEMP-no) or [Ba/Eu] $< 0$ (CEMP-r).
We excluded stars with high abundances of s-process elements (CEMP-s),
for which the most likely explanation seems to be pollution 
from a companion star.

\section{The chemical evolution model for the Milky Way halo}\label{model}

The chemical evolution model we use in this work is based on the
inhomogeneous model developed by Cescutti (2008) and  on
the homogeneous model of Chiappini et al. (2006).
We consider that the Galactic halo consists of 
100\footnote{We tested a larger number of volumes, and find that our
results converge after around 100 volumes, although increasing
considerably the computational time.} noninteracting cubic regions,
each with the same typical volume of $2.8\cdot10^{6}$ pc$^{3}$. In this
way the surface of the volume, taken as the surface of the side of a
cube, is $2\cdot10^{4}$ pc$^{2}$ ($S_{\mathrm{cube}}$).
The dimension of the volume is large enough to allow us to neglect the interaction
among different volumes  and small enough to
avoid losing the stochastic effects we are looking for. In fact, as we test for 
increasing volumes, the model tends to the homogeneous one (for details see Cescutti 2008).
We use time steps of 1Myr, which is shorter than any stellar lifetime considered here; in fact, 
the star with the maximum mass considered in the present work is 80 M$_{\odot}$, and its lifetime is $\sim$3 Myr.
At the same time, this time step is longer than the cooling time of the supernovae remnant bubbles, which is 
normally $\sim$ 0.1-0.2 Myr and 0.8 Myr at maximum  (for details see Cescutti 2008). 

We choose to use parameters of the chemical evolution
(star formation rate, initial mass function, stellar lifetime, nucleosynthesis)
that are very similar to those of the homogeneous model by Chiappini et al. (2006).
For this reason, the model is slightly different from the original
inhomogeneous model by Cescutti (2008), and it improves the agreement with
the metallicity distribution function of the Galactic halo (see Sect. 4).

In each region, we assume a Gaussian function as infall law of the gas with a 
primordial chemical composition:
\begin{equation}
\frac{d \scriptsize{\mathrm{Gas}}_{\scriptsize{\mathrm{Infall}}}(t)}{dt} =  A e^{\frac{-(t-t_{0})^{2}}{2 \sigma^{2}}}
\end{equation}
where the parameter $A$ is equal to $1.28\cdot10^{4} \mathrm{M}_{\odot}\mathrm{Myr}^{-1}$,
$t_{0}$ is 250 Myr, and $\sigma$ is 100 Myr.
As a result the formation of the halo takes place in 250 Myr.
In each cube we assume a star formation rate $\psi(t)$,
which is defined as:
\begin{equation}
\psi(t)= \nu~\Big(\frac{M_{\mathrm{gas}}(t)}{S_{\mathrm{cube}}}\Big)^{1.5}
\end{equation}
where $M_{\mathrm{gas}}(t)$ is the gas mass inside the considered box,
and the parameter $\nu$ is 0.6 M$_{\odot}^{-0.5} \mathrm{Myr}^{-1}$ pc $^{2}$.
We we assume gas outflows to be proportional to $\psi(t)$,
 following the prescriptions of Chiappini et al. (2006):
\begin{equation}
M_{\mathrm{out}}(t)=f \cdot \psi(t)
\end{equation}
where $f=14$.

Knowing the mass that is transformed into stars in a time step 
(hereafter, $M_{stars}^{\mathrm{new}}$), we assign the mass to one star with a random function,
 weighted according to an initial mass function (IMF).
In this work, we show the results obtained by the model 
by adopting two different IMFs:
\begin{itemize}
\item {\it Model 1} ``standard'' IMF of Scalo (1986) with two slopes:
\begin{equation}
\phi(M) = C\cdot M^{-(1+x)};\,
x = \left\{ \begin{array}{ll}
1.35 & \textrm{if $0.1~\mathrm{M}_{\odot}<M<~2.0~\mathrm{M}_{\odot}$}\\
1.70 & \textrm{if $2.0~\mathrm{M}_{\odot}<M<80.0~\mathrm{M}_{\odot}$}\\
\end{array} \right.\\
\end{equation}
\noindent
with~~$C=\frac{1}{\int^{80}_{0.1}M^{-x}~dM}$.
\vspace{0.15cm}
\\
\item {\it Model 2} ``top heavy IMF'' with the same slopes but 
with the mass range restricted between 1 and 80 M$_{\odot}$ 
when Z~$<~1~\cdot~10^{-6}$:
\begin{equation}
\phi(M) = C'\cdot M^{-(1+x)};\, x = 1.70\,\,\,\mathrm{if}\, 1.0~\mathrm{M}_{\odot}<M<80.0~\mathrm{M}_{\odot}
\end{equation}
\noindent
with~~C'=$ \frac{1}{\int^{80}_{1}M^{-x}~dM}$.
\vspace{0.15cm}
\\
For Z $> 1 \cdot 10^{-6}$  the normal mass range 
from 0.1 to 80 M$_{\odot}$ is applied,  assuming that around this 
critical metallicity the IMF switched to a normal one (see Schneider et al. 2002).
\end{itemize}
With the second IMF, we overcome the problem of having a high number
of long-living metal free stars.
Once the mass is assigned  to the first star, we extract the mass of 
another star, and repeat this cycle until
the total mass of newly-formed stars exceeds $M_{stars}^{new}$.
In this way, in each region, and at each time step, the $M_{stars}^{new}$ 
is the same, but the total number and mass distribution
 of the stars are different, and we know the mass of each star
contained in each region, when it is born, and when it will die, assuming
the stellar lifetimes of Maeder \& Meynet (1989).

We compute the chemical evolution in the following way:
at the end of its lifetime, each star enriches the interstellar medium
  with  newly-produced elements (see the next section) 
as a function of its mass and metallicity. 
The total mass of each element is determined at the end
of the lifetime of each star, by taking both the
newly formed and the preexisting chemical elements into account.

The model does not take the pollution produced by stars 
with mass $< 3 \mathrm{M}_{\odot}$ into account because their lifetimes exceed the 
duration of the simulation. 
The existence of SNe Ia is also taken into account, according to the prescriptions
 of Matteucci \& Greggio (1986), in the single degenerate scenario.
In fact, it is worth noting that the first Type Ia SNe occur after only 30~Myr, the lifetime
of an 8 M$_{\odot}$ star (see Matteucci \& Recchi 2001 for a discussion on this point).
With this model we are not only able to study the
impact of different sets of stellar yields in the chemical enrichment
of the Galactic halo, but also predict the expected spread for
different chemical elements at low metallicities, where the random
effects in the birth of stellar masses are important.

\begin{figure}
\begin{center}
\includegraphics[width=0.49\textwidth]{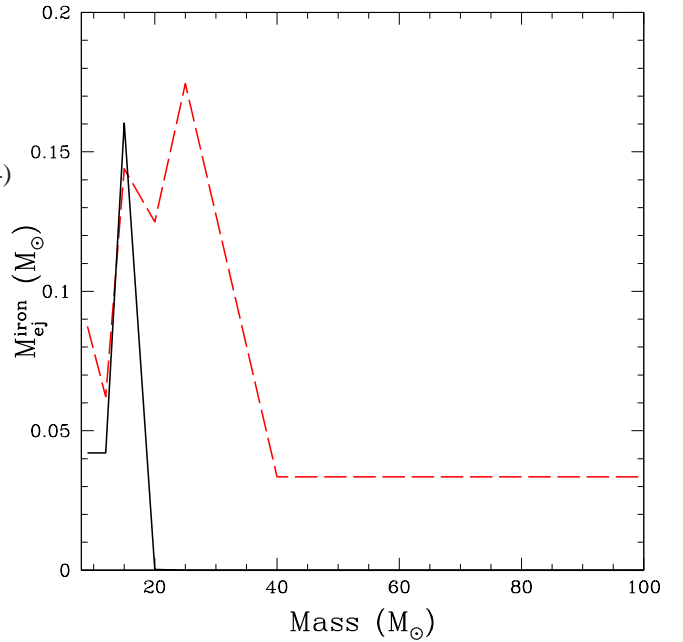}
\caption {Yields for iron computed by Limongi for $0<Z<10^{-8}$(solid line), 
compared to the yields computed by WW95 at solar metallicity (dashed line). }\label{yields}
\end{center}
\end{figure}

\section{Nucleosynthesis prescriptions}

In this section, we describe the set of the yields we adopt.
The chemical elements we investigate are CNO and Fe.
For the CNO yields, we use metal dependent yields for both low- and intermediate-mass 
stars and massive stars. For the low intermediate-mass stars, we adopt 
the stellar yields by Meynet \& Meader (2002) with rotation
for three metallicities, namely Z=0.02, Z=0.004, and Z=$10^{-5}$.
We assume that their table for Z=$10^{-5}$ is valid down to Z=0, 
as in Chiappini et al. (2003). The table for yields of
solar metallicity are published only in Chiappini et al. (2003).
For the massive stars, we again assume the set of the yields 
by Meynet \& Meader (2002) for the three metallicities described above.
We adopt two different sets of yields for Z=$10^{-8}$: 

\begin{itemize}

\item {\it Model a}: In the first set of yields, we consider the ejected chemical elements 
at the explosion of the star as SNII (as in Chiappini
et al. 2006), hereinafter named ``total yields'', because they are obtained
as the sum the contributions during both the stellar wind and
supernovae phases.

\item {\it Model b}: In the second set, we distinguish the yields  for stars more massive and less massive 
than 40 M$_{\odot}$. For the more massive stars, we only consider 
the chemical enrichment coming from the stellar wind, 
before the star collapses directly into a black hole (see Heger et al. 2003). 
We call this set of yields  ``wind yields''.  
For the stars less massive than 40 M$_{\odot}$ we again use the ejected chemical elements 
at the explosion of SNII.

\end{itemize}
The difference between total yields and wind yields is
only in the yields for  very massive stars ($M>40 \mathrm{M}_{\odot})$
and at very low metallicity, Z=10$^{-8}$.
Both the ``total'' yields and the ``wind'' yields
are taken from Hirschi (2007), when considering stars with very
high rotational velocity   (600 to 800 km/s).

 For iron, we adopt the solar metallicity yields of Woosley and Weaver (1995)
 for Z$>$ $10^{-8}$, whereas for 
Z $<$ $10^{-8}$ we adopt the yields of Limongi (private communication), and
the adopted yields of iron are shown in Fig. \ref{yields}.
These computations predict that population III stars with masses
above 20 M$_{\odot}$ inject negligible amounts of iron into the ISM upon their death.
This agrees with the idea that the more massive pristine stars will collapse directly 
into black holes (Heger et al. 2003).

\section{Results}

We test 4 different models (Models A,B, C, and D) with
different nucleosynthesis prescriptions and IMFs. The adopted
combinations of Models 1 and 2 (different IMFs) and Models a and b
(different set of stellar yields) are summarized in Table \ref{models}.

\begin{table}[ht!]

\centering
\begin{minipage}{85mm}
\caption{The nucleosynthesis prescriptions and IMFs adopted for the 4 models.}\label{models}
\begin{tabular}{|c|c|c|}
\hline

\hline\hline
Model&  IMF      &      yields \\
\hline 
A & ``top heavy'' (Model 2) & ``total yields'' (Model a)\\  
B & ``top heavy'' (Model 2)& ``wind yields'' (Model b)  \\ 
C & ``standard'' (Model 1) & ``total yields'' (Model a) \\
D & ``standard '' (Model 1) & ``wind yields''  (Model b) \\

\hline\hline

\end{tabular}

\end{minipage}

\end{table}

 In the figures that follow we only plot the abundance ratios of simulated stars 
still-living at the present time (i.e. stars with masses $<1 \mathrm{M}_{\odot}$).
\begin{figure}[ht!]
\begin{center}
\includegraphics[width=0.49\textwidth]{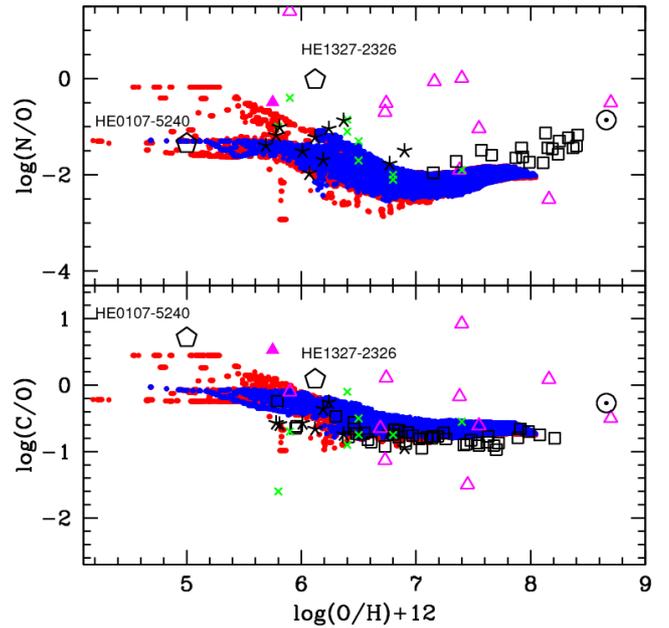}
\caption { log(N/O) vs $\log\epsilon$(O) (upper panel),
log(C/O) vs $\log\epsilon$(O) (lower panel); the still-living simulated stars 
from model A are in blue and from model B in red. Starred symbols are for the data from 
Spite et al. (2006),  crosses are for the data from Lai et al. (2008);
 the open square are data from Fabbian et al. (2009) for log(C/O)
and by Israelian et al. (2004) for log(N/O). 
The large open pentagons are the UMP stars, for which we insert the name of the star
close to the respective pentagon.
In this plot we also show the data for CEMP-no stars  with open triangles 
and for the CEMP-r star with a filled triangle. These data have been collected 
by Masseron et al. (2009).
The symbols of the Sun refer to the the solar values measured by Asplund et al. (2005).}
\label{fig1}
\end{center}
\end{figure}
In Fig. \ref{fig1} we show the results of model A (blue) and B (red) 
for the ratio of log(N/O) (top panel) and log(C/O ) (bottom panel).
Both models can reproduce the observed abundance trend for halo stars
in the range $6<\log\epsilon(\mathrm{O})<8$.
Our models do not extend to the metallicity range of the galactic disk, as they are pure halo models. 
The different models show different behaviors at very low metallicity. 
 At these earliest phases, in fact, the differences in the assumed yields are very important. 

In model B, we assume that the very massive stars end their lives directly as  black holes 
(see Heger et al. 2003), and only the newly produced elements  ejected by 
these massive stars in the stellar wind phase enrich the ISM, as explained in Sect.~3.
 The calculations of Hirschi  (2007) predict high N/O and C/O ratios in the stellar winds of
fast-rotating massive stars, higher than what is predicted for their total yields 
(i.e. including the material ejected during the SNII explosion).

In Fig. \ref{fig1} the stochasticity of the code clearly shows its
peculiarity, because it predicts that some of the stars observed today will
have  high C/O and N/O ratios (Model B), but also predicts that some stars
will have the same ratio as those predicted by Model A, which was
computed with total yields for the whole mass range. This is expected
because in Model B we adopted the wind yields only for the most massive
stars.  Such a result only can be obtained with an inhomogeneous
model, since standard homogenous chemical models would instead predict a
single trend representing the mean contribution of all the stars dying
at a given metallicity. This is a very interesting and important feature
of the chemical inhomogeneous model, and it represents the novelty of this approach.

The predictions of model B for N match the 
observed trend, and it is also able to explain the chemical abundances of the UMP
star HE~0107$-$5240 (see also Figs. 6 and 7 where the results for [C/Fe], [N/Fe], and 
[O/Fe] are shown). The spread predicted by model B is a consequence of our hypothesis that
massive stars have two different fates, with quite different 
nucleosynthesis at very low metallicities.
This assumption seems to be crucial for explaining the observed scatter in the (N/O) ratio of normal stars. 

However, this model is not able to explain the log(N/O) of HE~1327$-$2326, predicting
too low values for this ratio around 6.5 in $\log\epsilon(\mathrm{O})$. 
This suggests that HE~1327$-$2326 was essentially formed only from the 
wind material,  without dilution with the pristine ISM material. 
The formation from the pure wind material of new stars is not taken
 into account in our chemical model. In our model the ejecta from stars, whether
wind yields  or total yields, are mixed and diluted 
with the whole ISM gas present in the simulated region.

 In fact, this star shows low Li abundance, which is expected
if the material from which this object formed was essentially pure stellar wind 
material where the primordial Li was destroyed. In addition, this star has a potentially
low $^{12}$C/$^{13}$C (as only a lower limit of 5 is given by Aoki et al. 2006) typical 
of hydrogen-burned material (see Meynet et al. 2010). 
We notice that HE0107$-$5240 has a $^{12}$C/$^{13}$C =  60 $\pm$ 10 
(Christlieb, priv. communication), compatible with what is expected from 
an ISM enriched by fast-rotating stars (see Chiappini et al. 2008).

Model A, computed with ``total'' yields for all massive stars, does not reproduce the data as well as model B.
This is true not only for the UMP stars, but also for the EMPs, which show some objects with very high N/O ratios
towards the lowest values of $\log\epsilon(\mathrm{O})$, which is not predicted by the models.

For C/O, the measurements of metal-poor normal stars (Spite et al. 2006)
show much less spread than what is found for N/O. In fact, 
although stellar models with rotation also predict higher C/O,
this increase is much less important than the one in N. This means that the 
N yields are much more dependent on the initial rotational velocity, hence on the 
stellar mass, than the C yield. This explain the smaller scatter predicted for the C/O
than for the one of N/O.

\begin{figure}[ht!]
\begin{center}
\includegraphics[width=0.49\textwidth]{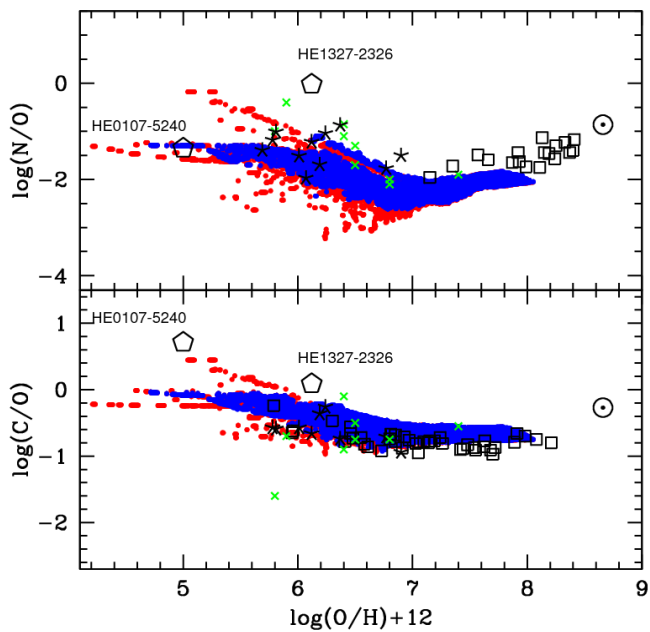}
\caption {log(N/O) vs $\log\epsilon$(O) (upper panel),
log(C/O) vs $\log\epsilon$(O) (lower panel). The still-living simulated stars 
from model C are in blue and from model D in red. Starred symbols are for the data from 
Spite et al. (2006),  crosses are for the data from Lai et al. (2008);
 the open square are data from Fabbian et al. (2009) for log(C/O),
and by Israelian et al. (2004) for log(N/O).  The large open 
pentagons are the UMP stars, for which we insert the name of the star
 close to the respective pentagon.
The symbols of the Sun refer to the the solar values measured by Asplund et al. (2005).}
\label{fig2}
\end{center}
\end{figure}

On the other hand, none of the models is able to reproduce the abundances of the 
UMP stars, although model B marginally agrees with the C/O of both
 HE~1327$-$2326 and HE~0107$-$5240.
 Very interestingly, none of our models is able to reproduce
the observed abundance ratios of the CEMP-no stars (see Fig. 2 for the C/O and N/O 
 and Fig. 6 for the C/Fe, N/Fe, and O/Fe). 
We thus conclude that the chemical abundances of the CEMP-no stars cannot be explained in the 
same framework as the abundances of normal VMP stars. The effect of stellar 
winds of fast rotators plus the inhomogeneity of the ISM are not strong enough to predict
the large observed enhancements in the CNO elements. As for the UMP stars discussed above
(also CNO enhanced), the CEMP-no stars were probably formed
 from almost pure wind material. It would be very useful to better constrain our scenario 
to have the abundances of Li and $^{13}$C for all CEMP-no stars (see Meynet et al. 2010).
The case of the only CEMP-r star for which we have the 
abundances of C, N, and O (represented by a filled triangle in Figs. 2 and 6)
is rather similar to the other CEMP-no discussed above.

In Fig. \ref{fig2} the predictions of models C and D (both models assume a standard IMF -- see table 1) 
are shown. The results are similar to the ones of models A and B. This suggests that it is not necessary 
to consider top-heavy IMF to explain the chemical abundances of CNO at low metallicity, once the 
contribution of fast rotators is taken into account.

\begin{figure}[ht!]
\begin{center}
\includegraphics[width=0.49\textwidth]{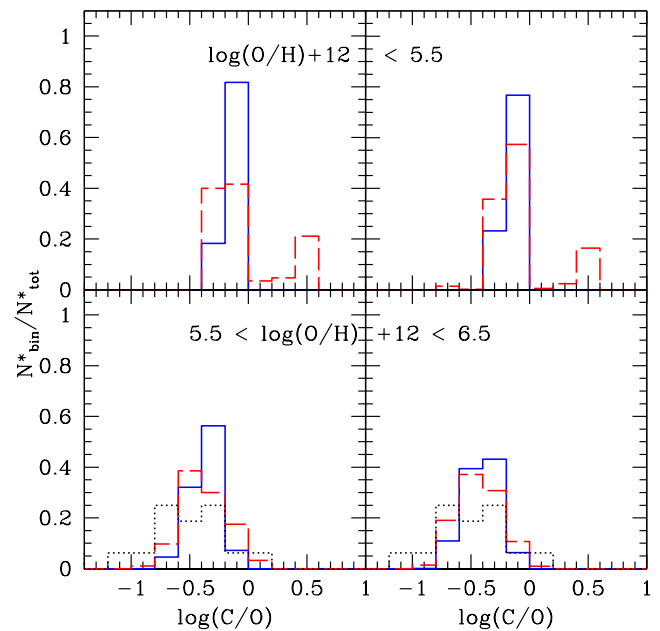}
\caption { In the left panels, the histograms for the distributions 
of simulated stars in bins of log(C/O), for models A (blue solid )
and B (red dashed). In the right panels, the histograms for the distributions 
of simulated stars in bins of log(C/O) for models C (blue solid )
and D (red dashed).  In the upper panels stars with $\log\epsilon(\mathrm{O}) < 5.5$, in the lower
panels stars with with $5.5 < \log\epsilon(\mathrm{O}) <  6.5$. In the lower panels,
the  distribution of the observed normal metal-poor is also shown  
stars with a  black dotted line.}
\label{fig8}
\end{center}
\end{figure}

\begin{figure}[ht!]
\begin{center}
\includegraphics[width=0.49\textwidth]{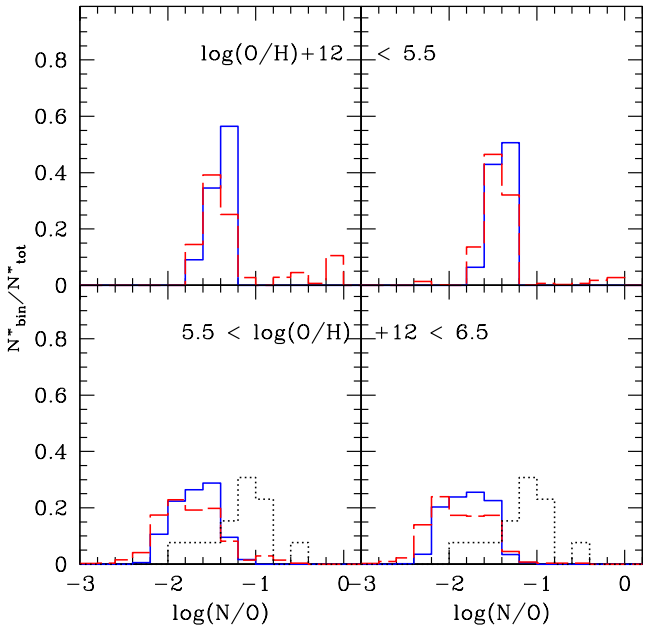}
\caption {  In the left panels, the histograms for the distributions 
of simulated stars in bins of log(N/O) for models A (blue solid )
and B (red dashed). In the right panels, the histograms for the distributions 
of simulated stars in bins of log(N/O) for models C (blue solid )
and D (red dashed). 
 In the upper panels we consider stars with $\log\epsilon(\mathrm{O}) < 5.5$, in the lower
panels stars with with $5.5 < \log\epsilon(\mathrm{O}) <  6.5$. 
In the lower panels, the distribution of the observed 
normal metal-poor stars is also shown with a black dotted line.}
\label{fig10}
\end{center}
\end{figure}

However, models A and B, with a top-heavy IMF, do overcome the problem
of producing a significative number of zero metallicity stars (still
living) which have not been observed up to now. In addition, models A and B
produce slightly more massive stars than models C and D,
as is barely visible when comparing Figs. \ref{fig1} and \ref{fig2}.
This can be seen more clearly in Fig. \ref{fig8}. This figure shows
the distribution of the simulated stars with respect to the C/O,
for the models A, B, C, and D.  To better disentangle the differences
at low metallicity we split the resulting stars in two ranges:
$\log\epsilon(\mathrm{O}) < 5.5$ in the upper panel, $ 5.5 <\log\epsilon(\mathrm{O}) < 6.5$ in
the lower panel.  In the upper panels, both models B (on the left) and
D (on the right) show a peak at $\log\mathrm{(C/O)}\sim 0$. On the right of this
peak, there is another lower peak.  This secondary peak is produced by
stars formed in a volume where the ISM was enriched by the stellar
winds of fast rotators, before fast rotators of lower mass have had time
to eject their products via the supernovae phase. As the stellar winds
of fast rotators are richer in C compared to what is ejected by the SN
explosion, these volumes will produce stars with higher C/O.
Most of the stars will in fact originate in a gas where the
contamination of less massive stars dominated (i.e. which polluted the
ISM with their total yields upon the explosion of the SN).

The second peak for model B is slightly higher than the one of model D 
because of the different IMF used (see Table \ref{models}), creating 
more massive stars. In the lower panels we compare the 
results of our models in a range of metallicity slightly broader,
 with the distribution of the observed stars in the same range
(we do not show the distribution of the observed stars in the upper 
panel because we only have the UMP star HE~0107$-$5240).
For C, in this metallicity range, the agreement between the observed distribution 
and what is predicted by model B is remarkable. In fact, our predictions agree
not only with the peak value but also with the observed spread.
 We underline that the number of stars observed in this range is very low (15)
and  larger statistics are needed to better constrain our models.
Model A reproduces the peak value too, but this model
does not predict the observed spread;  in particular, it does not produce
stars with high values for C/O. Toward a low value for this ratio, 
both model A and B  do not explain all the spread present in the observed
stars, but again the model B predicts  this feature slightly better.
The results for models C and D are very similar to the ones of models A and
B, since the top-heavy IMF does not influence much at this stage, but model B 
still seems to reproduce the spread in the ratio of C/O for 
the observed stars slightly better than model D.

Our predictions for the distribution of N/O, for the four models are shown in 
Fig. \ref{fig10}. In the top panel, where the results are shown for $\log\epsilon(\mathrm{O}) < 5.5$,
models computed with wind yields do not produce a secondary peak as for carbon,
but  rather a flat distribution towards higher log(N/O) values. This is due to 
the different behavior of adopted yields for very massive stars in the 
set of the wind yields. Nevertheless, again  model B compared to model D
shows higher number of stars with an enhanced log(N/O), again
due to the different IMF adopted.

 In the lower panel  for $ 5.5 <\log\epsilon(\mathrm{O}) < 6.5$, we show the distribution for the observed stars, too.
On the bottom left, it can be see that the predicted N/O by model B shows a spread that is very similar to the 
observed one but that predicts a ratio around 0.4~dex lower than the one indicated by the data.
In the next figures, we show that the reason for this
difference relies on the yields of N (rather than O) in the $ 5.5 <\log\epsilon(\mathrm{O}) < 6.5$ metallicity range. 
Model A presents a narrower spread for N/O than model B. 
 Models C and D (bottom right) show similar results to models A and B, respectively, at this metallicity range.

\begin{figure}[ht!]
\begin{center}
\includegraphics[width=0.49\textwidth]{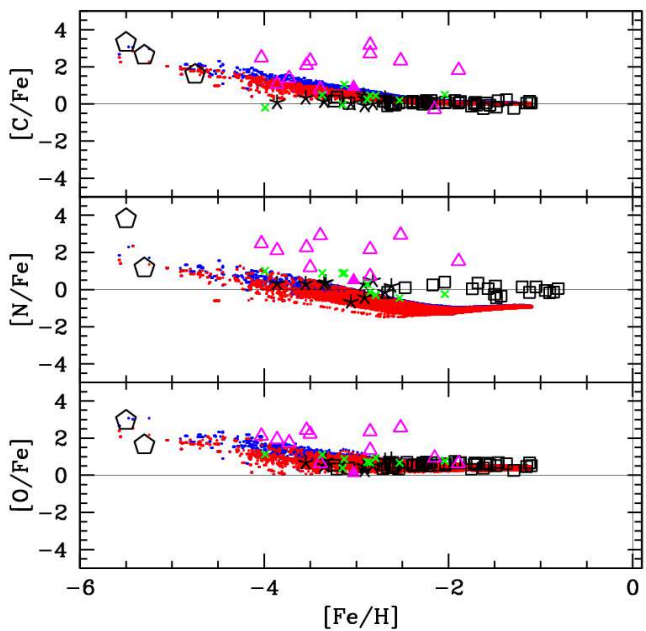}
\caption {[C/Fe] vs [Fe/H] (upper panel),
[N/Fe] vs [Fe/H] (middle panel),
[O/Fe] vs [Fe/H] (lower panel).
The still-living simulated stars for model A in blue and for model B in red, 
starred symbols are for the data from Spite et al. (2006), 
crosses are the data from Lai et al. (2008). 
The open squares are data from Israelian (2004) for [N/Fe]
and from Fabbian et al. (2009) for [C/Fe] and [O/Fe].
The large pentagons are the UMP, described in the observational data.
 In this plot we also show the data for CEMP-no stars  with open triangles 
and for the CEMP-r star with  a filled triangle. These data have been collected 
by Masseron et al. (2009). }\label{fig3}
\end{center}
\end{figure}

\begin{figure}[ht!]
\begin{center}
\includegraphics[width=0.49\textwidth]{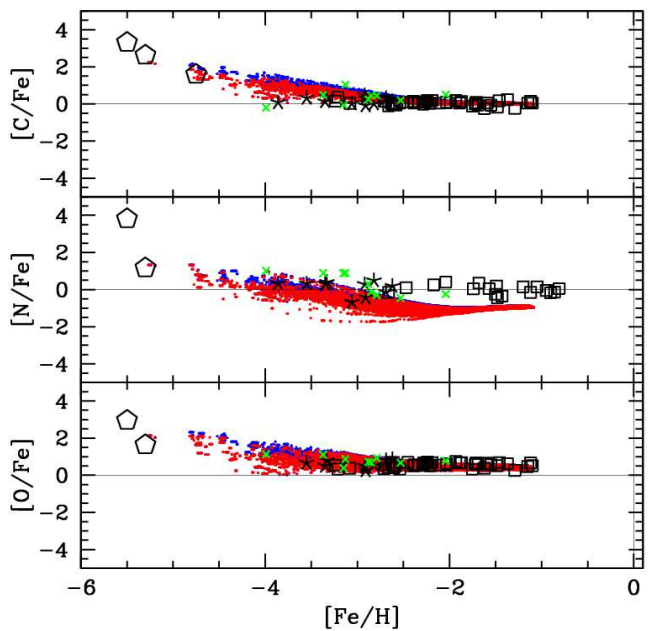}
\caption { [C/Fe] vs [Fe/H] (upper panel),
[N/Fe] vs [Fe/H] (middle panel),
[O/Fe] vs [Fe/H] (lower panel).
The still-living simulated stars for model A in blue and for model B in red, 
starred symbols are for the data from Spite et al. (2006), 
crosses are the data from Lai et al. (2008). 
The open squares are data from Israelian (2004) for [N/Fe]
and from Fabbian et al. (2009) for [C/Fe] and [O/Fe].
The large pentagons are the UMP, described in the observational data.}
\label{fig4}
\end{center}
\end{figure}

The results presented in Figs. \ref{fig8} and \ref{fig10} illustrate that inhomogeneous
models can go beyond the interpretation of abundance ratio trends with
metallicity, also predicting scatter at the different phases of the
chemical enrichment history of the Galactic halo. At present, the poor
statistics of the observations limit the utility of such
models. Here a comparison between our theoretical models and
observations is only possible in a very limited metallicity range --
$ 5.5 <\log\epsilon(\mathrm{O}) < 6.5$ --, due to the lack of data.
 This situation will improve soon with the upcoming large datasets for halo stars (from
surveys such as SEGUE-2 and LAMOST -- see Beers 2010), which will play
a fundamental role in constraining this kind of model.

Our predictions  for [C/Fe], [N/Fe], and [O/Fe] are shown in Figs. \ref{fig3} (models
 A and B) and \ref{fig4} (models C and D). In these plots we  consider
the solar abundances by Asplund et al. (2005).
The results of the models in these plots are not completely
self-consistent from the point of view of nucleosynthesis as the iron
yields came from different authors who include the explosive nucleosynthesis, whereas the CNO yields
come from the theoretical predictions of the Geneva group, which include only the pre-supernovae phase.
Our models can reproduce these ratios well, and
also predict some stars in the locus of HE~0107$-$5240, whereas HE~1327$-$2326 is not reproduced.
Moreover, our models cannot account for the CEMP-no stars. For nitrogen, there is a marginal
disagreement in the intermediate metallicity range, pointing to an extra source of nitrogen
not accounted for in our models. As discussed in Chiappini et al. (2006, 2008) and Pipino et al. (2009),
this is most probably because current chemical evolution models do not include 
the contribution of super AGB stars. Super AGB stars could contribute with important quantities
of nitrogen, and their contribution would be visible just before the appearance of the 
intermediate-mass stars (Chiappini et al. in prep.).

Finally, although our models cannot explain the [N/Fe] ratio of HE~1327$-$2326, 
we are able to reach the very low enrichment 
in Fe observed in this star and to predict almost the correct ratios of [C/Fe] and [O/Fe]
in models A and B, in the chemical inhomogeneous framework. 
Moreover, we reproduce the ratio of the three elements [C/Fe], [N/Fe], and [O/Fe] 
for the other star with an [Fe/H] $\lesssim -5$, namely, HE~0107$-$5240.
We also reproduce HE~0557$-$4840 discovered and analyzed by Norris et al (2007)
for which we only have the [C/Fe] ratio.

\section{Conclusions}

In this paper we have shown the results of an
inhomogeneous chemical evolution model for the
mean trend and scatter of C/O and N/O in the Galactic halo,
taking the contribution of fast-rotator stars into account.

The observational studies of chemical enrichment in
very metal-poor stars of the halo have found a large scatter 
(larger than the uncertainties in the derived abundances)
for the measured C/O and N/O. 
A large scatter was also found for neutron capture 
 elements in the same stars. These findings contrast
with the results for $\alpha$-elements, which instead presented striking
homogeneous [$\alpha$/Fe] ratios. 
In Cescutti (2008), the spread in the chemical abundance ratios of 
neutron capture elements has been explained with a different
mass range for the production of these elements (from 12 to 30 M$_{\odot}$),
compared to the whole range of massive stars for the $\alpha$-elements.

In this work the nucleosynthesis differences for the elements C, N, and O  
come from the rotation of massive stars, which strongly affects the ratio of the production
among these elements at low metallicity (see Hirschi et al. 2007).
Moreover, we consider two possible contributions to the enrichment
of the ISM by  massive  stars at very low metallicity, the usual
enrichment through supernovae ejecta, and the enrichment
 only through their stellar winds.

 We find that the assumption that the most massive fast rotators
only contribute to the ISM enrichment via stellar winds leads to chemical evolution models that are able to
 account for both the large scatter in the N/O and C/O and the simultaneous lack of scatter 
in $\alpha$-elements observed in very metal-poor normal halo
stars. 
In fact, the stellar yields of
the latter elements do not show any strong dependency on 
the stellar mass, contrary to what happens for N, C, and n-capture
elements.

In this context, we also explored whether the scatter in N and C created by
the strong yields dependence on the stellar mass, when including
the fast rotators, would also account for the existence of the
so-called carbon-enhanced stars (in particular the CEMP-no ones). We
find that, even when considering that the more massive fast rotators
enrich the early ISM mainly via stellar winds strongly enhanced in
CNO, it is impossible to explain the abundances observed in CEMP-no
stars. However, we point out that the latter model  marginally
agrees with the abundance ratios of the only CEMP-r star for which
the CNO abundances have been measured, and of the two of the three
UMP stars known to date, HE~0107$-$5240 and HE~0557$-$4840. These results
suggest that very metal-poor massive stars would collapse directly
into black holes as predicted by Heger et al. (2003).

The discrepancy with respect to the CEMP-no stars, and in particular, HE~1327$-$2326, 
suggest that these objects are born from the gas expelled during the wind phase of
fast rotators, without much mixing with the surrounding ISM. A way to
test this hypothesis is to look for the abundances of $^{7}$Li, $^{12}$C/$^{13}$C,
 and helium in CEMP-no stars. In fact, the wind material is depleted in $^{7}$Li
and enriched in $^{13}$C (see Meynet et al. 2010 and Chiappini et al. 2008).

This work illustrates the importance of future surveys that will
enlarge the samples of very metal-poor stars with good abundance
determinations. The large statistics brought by these large datasets
will enable  detailed study not only of the abundance trends but
also of the intrinsic scatter in those trends and its metallicity
dependence. On the other hand, it is clear that high-quality abundances
for key elements such as CNO, the C isotopic ratio, n-capture elements,
and $^{7}$Li for a large sample of stars (including CEMP-no stars) is still
needed. Spectroscopic follow-ups of metal-poor stars found in the
ongoing and planned large surveys (e.g. SEGUE-2 and LAMOST) will play
a crucial role in constraining the stellar nucleosynthesis of the
first generation of massive stars and in unveiling the role of the fast
rotators in the early Universe.

\begin{acknowledgements}
G.C. acknowledges financial support from the Fondazione Cassa di Risparmio di Trieste.
G.C. and C.C. acknowledge financial support from PRIN2007-MIUR (Italian Ministry of University and Research), 
Prot.2007JJC53X-001.
G.C. would like to thank F. Matteucci for useful discussions and suggestions.
C.C. acknowledges financial support from the Swiss National Foundation (SNF).
C. C. thanks  R. Hirschi, S. Ekstr\"om, G.Meynet, and M. Limongi for providing 
the stellar yields and for fruitful discussions.

\end{acknowledgements}

\end{document}